# Integrated Supermode Photonics Enabled by Supersymmetric Transformation

Kaile Chen[1], Qi Lu[1], Yuan Zhong[1], Jingchi Li[1], Yuru Li[2], Zhaohui Li[2, 3], Chao Lu[4], Yikai Su[1*], and Lu Sun[1*]

[1]State Key Lab of Photonics and Communications, Department of Electronic Engineering, Shanghai Jiao Tong University, Shanghai 200240, China

[2]School of Microelectronics Science and Technology, Sun Yat-Sen University, Zhuhai 519000, China

[3]Southern Marine Science and Engineering Guangdong Laboratory (Zhuhai), Zhuhai 519000, China

[4]Photonics Research Institute, Department of Electronic and Information Engineering, The Hong Kong Polytechnic University, Hong Kong SAR, China

[*]Corresponding author: Yikai Su and Lu Sun (email: yikaisu@sjtu.edu.cn; sunlu@sjtu.edu.cn)

## Abstract

As a fundamental degree of freedom of optical fields, mode plays a central role in emerging applications such as artificial intelligence, optical computing and quantum information processing. However, conventional multimode waveguides feature non-equidistant effective index distribution for eigenmodes, which makes mode orthogonality vulnerable to perturbations, especially when two modes are close in effective index. It limits the fidelity and scalability of mode channels for applications involving multiple modes such as mode-division multiplexing and mode-encoded high-dimensional entanglement. To circumvent these limitations, supermode photonics, which leverages highly engineerable supermodes in coupled waveguides, has emerged as an alternative approach for controlling multiple mode channels simultaneously and achieving high parallelism in information processing. Nevertheless, precise supermode excitation and intermodal crosstalk suppression remain challenging for the practical deployment of supermode photonics. Here, we report a systematic methodology to obtain supermodes with equidistant effective index distribution and to excite arbitrary target supermodes with high precision. By employing a multi-well optical potential realized by a judiciously designed waveguide array, the supported supermodes achieve maximal spacing and an equidistant distribution in effective index, thereby efficiently suppressing the intermodal crosstalk during propagation that would otherwise arise from the closely spaced modes supported by the single-well optical potentials of conventional multimode waveguides. More importantly, we develop a 2nd-order discrete supersymmetric (DSUSY) transformation method that enables the excitation and detection of two supermodes at the same time and can be extended to any number of supermodes via simple cascading. Together, these findings overcome the long-standing bottlenecks in integrated supermode photonics and provide an intrinsically scalable route towards harnessing supermodes as a new degree of freedom for encoding, transmitting, and processing information. We experimentally demonstrate the feasibility and universality of this method by realizing two- and four-supermode multiplexing systems. Benefitting from the large effective index spacing between supermodes and the isospectral nature of the DSUSY transformation, the fabricated devices show low insertion losses (< 2.48 dB at 1550 nm) and intermodal crosstalk (< -18 dB at 1550 nm) for all mode channels over a 100-nm wavelength range (1500-1600 nm). The high-speed data transmission experiment performed on the four-channel system achieves an aggregate data rate of 1.024 Tb/s while maintaining considerably low bit error rates, underscoring the potential of supermode photonics for high-capacity on-chip optical communications. This work lays the foundation for integrated supermode photonics, which uses supermodes as a new degree of freedom for light manipulation and opens new avenues for supermode-based applications including but not limited to on-chip optical communications, intelligent optical computing and quantum information technologies.

## Introduction

The rapid growth of 5G/6G communications, cloud computing and artificial intelligence has driven great efforts to fully exploit the degrees of freedom of light. Mode, as an important optical dimension, has attracted extensive research interest and given rise to the field of multimode photonics[1-8]. However, conventional multimode waveguides feature single-well optical potentials, which inherently lead to non-equidistant distributions of effective modal indices ($n_{eff}$)[9]. As a result, even slight fabrication imperfections can induce severe intermodal crosstalk (CT) between two waveguide modes that are closely spaced in $n_{eff}$[10,11], hindering the development of large-scale integration and high-yield mass production of multimode photonic devices[12,13].

Recently, supermode photonics has emerged as a promising solution to realizing mutually orthogonal modal channels while suppressing inter-channel CT under perturbations[14,15]. Different from conventional multimode waveguides, the coupled waveguides supporting supermodes can be described by multi-well optical potentials, enabling highly tailorable $n_{eff}$ distributions for supermodes[9,16,17]. Here, we first propose an energy level engineering approach to achieving an equidistant $n_{eff}$ distribution for supermodes. By tuning the waveguide widths and gaps in the coupled-waveguide array, one can realize a multi-well potential that supports supermodes with uniform and maximized $n_{eff}$ spacing. This feature enables the robust manipulation of multiple supermodes at the same time, ensuring low insertion losses (ILs) and minimal CT even in the presence of fabrication imperfections.

However, it requires precise control over the relative intensities and phases of the optical fields in individual waveguides to achieve high-purity supermode excitation in the coupled-waveguide array, which is extremely challenging at the subwavelength scale[18,19]. To address this issue, supersymmetric (SUSY) transformation, originally developed in quantum mechanics[20,21], has been introduced to integrated photonics as a powerful tool for supermode manipulation[9,17]. In the past few years, SUSY transformations have found wide applications in designing diverse photonic devices, ranging from high-power single-mode laser arrays[22-25], to waveguide mode converters and multiplexers with continuously varying refractive indices[18,19,26-29], and perfect excitation of edge states in topological waveguide arrays[30-32]. These pioneering works clearly demonstrate the isospectrality of SUSY transformations when engineering the optical potentials of waveguide arrays[33]. However, they have focused on manipulating a single supermode using SUSY transformations, and the simultaneous control of multiple supermodes remains elusive. More recently, supermode-division multiplexing has been realized in densely packed waveguide arrays[34,35] and multi-core bus waveguides[36], where supermodes are mostly localized in individual waveguides with large waveguide width difference. This compromises the aforementioned highly tailorable $n_{\text{eff}}$ distributions of supermodes and requires extra operations such as mode exchange to implement supermode (de)multiplexing. It limits the scalability of this architecture to systems with larger channel numbers as it demands customized geometric optimization for each supermode rather than tackling multiple supermodes at the same time using SUSY transformations. Therefore, how to simultaneously manipulate multiple supermodes in a scalable manner remains an open challenge.

Here, we propose and experimentally demonstrate a 2nd-order discrete SUSY (DSUSY) transformation method that can, in principle, handle an arbitrary number of supermodes. By adiabatically connecting the original waveguide array with its SUSY partner, two target supermodes can be simultaneously launched or extracted with high fidelity through two isolated waveguides. This method can be straightforwardly extended to arbitrarily large-scale supermode multiplexing by cascading multiple SUSY structures. As proofs of concept, we implement two- and four-supermode multiplexing systems with low ILs (< 2.48 dB at 1550 nm) and intermodal CT (< -18 dB at 1550 nm) for all channels in the wavelength range of 1500-1600 nm. A high-speed data transmission experiment was also conducted on the four-channel system, achieving an aggregate data rate of 1.024 Tb/s with bit error rates (BERs) well below the KP4 forward error correction (KP4-FEC) threshold. This work reveals the potential of supermode photonics for high-capacity on-chip optical communications[37,38], and provides a paradigm for utilizing supermodes as a new degree of freedom in many advanced applications, such as large-scale optical computing[2,6] and high-fidelity quantum information processing[39,40].

## Results

### Effective index engineering and SUSY transformation of supermodes

Figure 1a shows the cross sections of three types of waveguide structures (top) and their corresponding optical potentials and energy level distributions (bottom). The effective Hamiltonians of the multimode waveguide, the supermode waveguide array and the DSUSY-transformed waveguide array are denoted as $H_{\mathrm{MM}}$, $H_{\mathrm{SM}}$ and $H_{\mathrm{DSUSY}}$, respectively. The traditional multimode waveguide can be described by a single square well potential, which supports eigenmodes with non-equidistantly distributed $n_{\mathrm{eff}}$. It will cause severe intermodal CT in the presence of fabrication imperfections when two modes are close in $n_{\mathrm{eff}}$ (e. g., the modes corresponding to the red and yellow energy levels of the single-well potential in Fig. 1a). To address

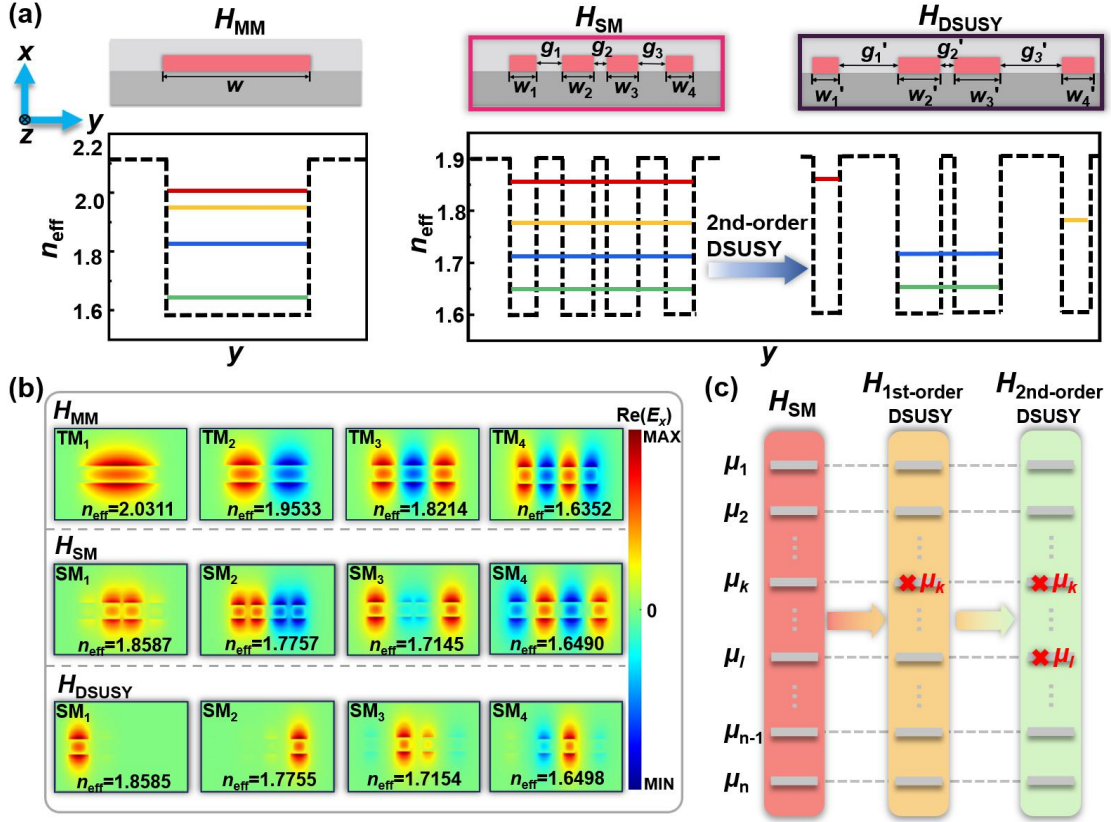

**Fig. 1| Principle of supermode photonics based on SUSY transformation. a,** Cross sections (top) and effective optical potentials (bottom) of three waveguide structures: a conventional multimode waveguide (left), a supermode waveguide array (middle) and a DSUSY-transformed waveguide array (right). The black dashed and colorful solid lines indicate the optical potentials and the energy levels, respectively. **b,** Simulated mode profiles ( $\text{Re}(E_x)$ ) and effective indices ( $n_{eff}$ ) for the eigenmodes supported by the waveguide structures with effective Hamiltonians $H_{MM}$, $H_{SM}$ and $H_{DSUSY}$. **c,** Schematic description of the 2nd-order DSUSY transformation. Two DSUSY transformations are performed consecutively on $H_{SM}$ to derive the superpartner $H_{\text{2nd-order DSUSY}}$, with two eigenvalues $\mu_i$ and $\mu_j$ eliminated from the spectrum and the other eigenvalues remained intact.

this issue, we engineer the structure of the supermode waveguide array to realize a multi-well optical potential, where the widths of the potential wells and barriers are determined by the waveguide widths $w_i$ and the gaps between waveguides $g_j$. To achieve an equidistant $n_{eff}$ distribution with maximized spacing, we choose the following parameters for the waveguide array implemented on the silicon-on-insulator (SOI) platform: $w_1 = w_4 = 437$ nm, $w_2 = w_3 = 516$ nm, $g_1 = g_3 = 247$ nm and $g_2 = 152$ nm. The detailed design and optimization procedure can be found in Supplementary

Note 1. The simulated profiles ($\text{Re}(E_x)$) and effective indices ($n_{\text{eff}}$) of the transverse magnetic (TM) modes supported by the multimode waveguide and the supermode waveguide array are compared in Fig. 1b. Obviously, the multimode waveguide exhibits a non-equidistant $n_{\text{eff}}$ distribution, as compared to the nearly uniform $n_{\text{eff}}$ spacing in the supermode waveguide array.

To achieve high-purity excitation of supermodes in the supermode waveguide array, we employ a 2nd-order DSUSY transformation to obtain the superpartner structure with isolated waveguides. The dynamics of light propagation in the supermode waveguide array is governed by the effective Hamiltonian $H_{\text{SM}}$[41], where the diagonal elements of $H_{\text{SM}}$ represent the propagation constants $\beta_i$ in individual waveguides, and the off-diagonal elements correspond to the nearest-neighbor coupling coefficients $\kappa_{i,i+1}$ between waveguides. A more detailed derivation of $H_{\text{SM}}$ is provided in Supplementary Note 2. Then, two successive DSUSY transformations are applied to $H_{\text{SM}}$ to construct the superpartner $H_{\text{DSUSY}}$, as illustrated in Fig. 1c. In the process, one can selectively remove two eigenvalues $\mu_k$ and $\mu_l$ from the spectrum of $H_{\text{DSUSY}}$ while preserving the remaining eigenvalues[30,31,42], which corresponds to decoupling two target supermodes from the evanescently coupled waveguides. This procedure can be mathematically described as follows:

$$H_{\text{SM}} - \mu_k I = Q_1 R_1, \tag{1a}$$

$$H_{\text{1st-order DSUSY}} = R_1 Q_1 + \mu_k I, \tag{1b}$$

$$H_{\text{1st-order DSUSY}} - \mu_l I = Q_2 R_2, \tag{1c}$$

$$H_{\text{2nd-order DSUSY}} = R_2 Q_2 + \mu_l I, \tag{1d}$$

where $I$ denotes the identity matrix, and $Q_{1,2}$ and $R_{1,2}$ represent the orthogonal and upper triangular matrices of QR factorizations, respectively. The explicit expressions for $H_{\text{SM}}$ and $H_{\text{DSUSY}}$ used in this work are detailed in Supplementary Note 2. Here, we first eliminate the effective indices associated with the SM$_1$ and SM$_2$ modes and obtain the 2nd-order superpartner $H_{\text{DSUSY}}$. Following the relationships between the matrix elements of $H_{\text{DSUSY}}$ and the geometric parameters of the waveguide array (see Supplementary Note 2 for details), we implement the effective Hamiltonian with a waveguide lattice model. The structure, the corresponding optical potential and the energy level distribution of the DSUSY-transformed waveguide array are shown in Fig. 1a (rightmost), with the simulated supermode profiles ($\text{Re}(E_x)$) and effective indices ($n_{\text{eff}}$) displayed in Fig. 1b (bottom). The structural parameters of the waveguide array are $w'_1 = 655$ nm, $w'_2 = 410$ nm, $w'_3 = 375$ nm, $w'_4 = 504$ nm, $g'_1 = 480$ nm, $g'_2 = 455$ nm and $g'_3 = 530$ nm. The gaps $g'_1$ and $g'_3$ are large enough to isolate the two outer waveguides from the central coupled-waveguide array. The isospectrality between the supermodes supported by the original waveguide array and the transformed counterpart can be clearly observed in these figures. By adiabatically connecting the waveguide arrays before and after the DSUSY transformation, the TM$_0$ modes launched into the two isolated outer waveguides of the transformed waveguide array (represented by $H_{\text{DSUSY}}$) will excite the target SM$_1$ and SM$_2$ supermodes in the original waveguide array (represented by $H_{\text{SM}}$). It offers a scalable scheme for high-fidelity excitation and extraction of supermodes in coupled waveguides, as will be detailed below.

Similarly, we can eliminate the effective indices associated with the SM$_3$ and SM$_4$ modes and apply a 2nd-order DSUSY transformation to $H_{\text{SM}}$ to obtain another superpartner $H'_{\text{DSUSY}}$. Again, the effective Hamiltonian $H'_{\text{DSUSY}}$ can be implemented by constructing a new waveguide array with the following parameters: $w''_1 = 424$ nm, $w''_2 = 600$ nm, $w''_3 = 547$ nm, $w''_4 = 352$ nm, $g''_1 = 650$ nm, $g''_2 = 218$ nm and $g''_3 = 650$ nm (see Supplementary Note 2 for details). In this case, the TM$_0$ modes supported by the isolated waveguides of the DSUSY-transformed waveguide array evolve adiabatically to the target SM$_3$ and SM$_4$ supermodes of the original waveguide array.

## Two-supermode multiplexing based on SUSY transformation

To prove the feasibility and universality of the proposed method in controlling multiple supermodes with an equidistant $n_{\text{eff}}$ distribution, we design two supermode-division multiplexing devices using the above-mentioned 2nd-order DSUSY transformation for the SM$_1$ and SM$_2$ modes and the SM$_3$ and SM$_4$ modes, respectively. For example, the SM$_1$ and SM$_2$ mode (de)multiplexer is composed of a pair of mirrored adiabatic evolution regions. One of them is encircled by the orange dashed box and enlarged on the right, where the structure is gradually transformed from the waveguide lattice described by $H_{\text{DSUSY}}$ to that described by $H_{\text{SM}}$ (Fig. 2a). The geometry of the transition region is defined by the following functions:

$$w_i(z) = m(z)w_i + (1-m(z))w_i', \quad i=1, 2, 3, 4 \tag{2a}$$

$$g_j(z) = m(z)g_j + (1-m(z))g_j', \quad j=1, 2, 3 \tag{2b}$$

where $z$ is the propagation distance, and $w_i(z)$ and $g_j(z)$ represent the variation functions of the waveguide widths and inter-waveguide gaps along the propagation direction. $w_i$ ($w_i'$) and $g_j$ ($g_j'$) correspond to the waveguide widths and inter-waveguide gaps for $H_{\text{SM}}$ ($H_{\text{DSUSY}}$), respectively. $m(z)$ can be any slowly varying function satisfying $m(z) \in [0,1]$ and $m(0)=0$, $m(L)=1$, where $L$ is the length of the transition region. For the sake of simplicity, here we choose a linear variation function $m(z) = \frac{z}{L}$. The SM$_3$ and SM$_4$ mode (de)multiplexer can be constructed using a similar structure (implementing the superpartner $H'_{\text{DSUSY}}$ instead of $H_{\text{DSUSY}}$). To guarantee the adiabaticity of the evolution[43], the length $L$ is set to be 200 μm(160 μm) for the SM$_1$ and SM$_2$(SM$_3$ and SM$_4$) mode (de)multiplexing. It can be further reduced by employing various strategies, including shortcuts to adiabaticity (STA)[44], quantum metric engineering[45], effective

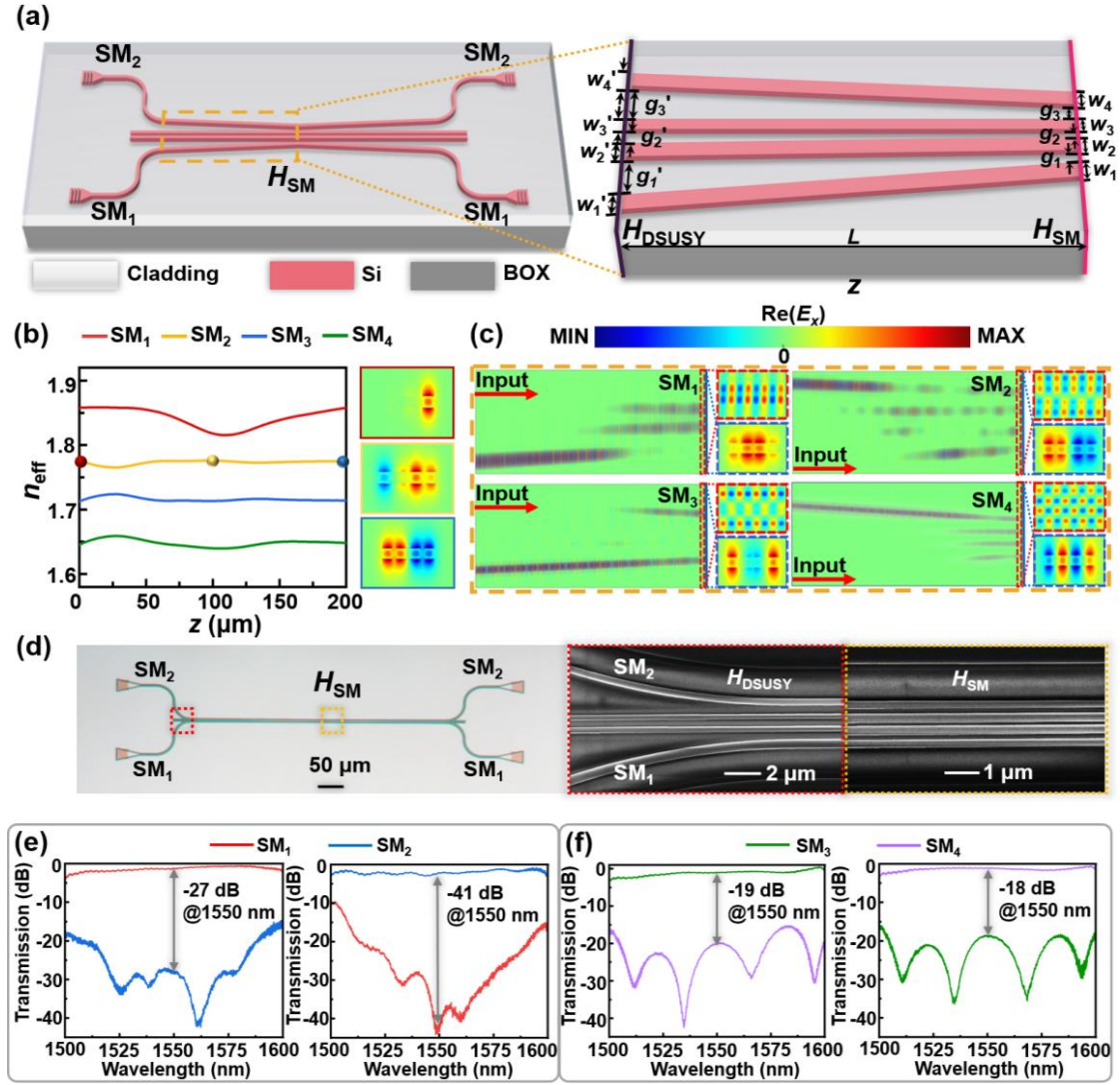

**Fig. 2| Two-supermode multiplexing systems enabled by 2nd-order DSUSY transformations.
a,** Schematic of the proposed two-supermode multiplexing device (left) and an enlarged view of the adiabatic evolution region encircled by the orange dashed box (right). The DSUSY waveguide array (represented by $H_{\text{DSUSY}}$) is adiabatically transformed into the supermode array (represented by $H_{\text{SM}}$) in the transition region. **b,** Evolution of the effective indices $n_{\text{eff}}$ of the supermodes along the propagation direction $z$. The insets on the right display the mode profiles ($\text{Re}(E_x)$) of the SM2 supermode at $z = 0, 100, 200\ \mu m$, which correspond to the red, yellow and blue dots on the evolution trajectory of $n_{\text{eff}}$, respectively. **c,** Simulated propagation profiles ($\text{Re}(E_x)$) in the adiabatic evolution region when the light is launched from the isolated waveguides. The insets on the right show the top-view (in the red dashed boxes) and cross-sectional (in the blue dashed boxes) electric field distributions at the output of the waveguide array. **d,** Optical microscope photo of the fabricated device (left) and SEM images of the areas of the waveguide array encircled by the red and orange dashed boxes (right). **e, f,** Measured transmission spectra of the fabricated two-supermode multiplexing systems for (**e**) the SM1 and SM2 channels and (**f**) the SM3 and SM4 channels, respectively.

Berry connection minimization[46], adiabatic passage[47], and inverse design methods[48]. Then, we use a commercial eigenmode solver (Lumerical MODE Solutions) to calculate the effective indices $n_{eff}$ of all supermodes along the propagation direction of the $SM_1$ and $SM_2$ mode multiplexing region. As shown in Fig. 2b, the four supermodes remain well separated in $n_{eff}$ throughout the adiabatic evolution, successfully avoiding mode crossings and therefore suppressing intermodal CT under perturbations. The insets on the right present the mode profiles ($Re(E_x)$) of the $SM_2$ supermode at three representative positions ($z$ = 0, 100, 200 μm), corresponding to the red, yellow and blue dots on the evolution trajectory of $n_{eff}$. It suggests that the $TM_0$ mode localized in the isolated waveguide of the DSUSY lattice will be adiabatically converted to the $SM_2$ supermode in the coupled waveguides of the supermode lattice during propagation. The quasi-isospectrality of the supermodes along the propagation direction is also verified by the almost flat trajectories of $n_{eff}$. This property constitutes the key mechanism underlying the precise excitation and extraction of supermodes in our design[31,32].

Next, three-dimensional finite-difference time-domain (3D FDTD) methods were employed to simulate the adiabatic evolution of supermodes in the transition region. Figure 2c displays the propagation profiles ($Re(E_x)$) in this region when the light is launched from the isolated waveguides of the DSUSY array. For the $SM_1$ and $SM_2$ mode multiplexer, the $TM_0$ mode injected from the bottom(top) waveguide adiabatically evolves into the $SM_1$($SM_2$) supermode at the output facet. In the $SM_3$ and $SM_4$ mode multiplexing case, the input from the bottom(top) waveguide yields the excitation of the $SM_3$($SM_4$) supermode. The insets next to the propagation profiles exhibit the enlarged top-view (in the red dashed boxes) and cross-sectional (in the blue dashed boxes) electric field distributions ($Re(E_x)$) at the output of the waveguide array. It shows clear evidence that the target supermodes are successfully excited with high mode purity. The simulated transmission spectra of these supermodes were also extracted to quantitively characterize the performance of two-supermode (de)multiplexers (see Supplementary Note 3 for details). Over a broad bandwidth covering 1500-1600 nm, the ILs are less than 0.15 and 0.22 dB and the CT values are below -15.4 and -16 dB for the $SM_1$ and $SM_2$ channels, respectively (ILs of < 0.05 dB and CT of < -21 dB for both channels at 1550 nm). In the same wavelength range, the $SM_3$ and $SM_4$ mode (de)multiplexer exhibits low ILs of < 0.13 and < 0.06 dB and CT of < -18.4 and < -18.7 dB for the $SM_3$ and $SM_4$ channels, respectively (ILs of < 0.11 dB and CT of < -21 dB for both channels at 1550 nm). Together, these numerical results confirm the broadband, low-loss and low-CT performance of two-supermode multiplexing enabled by the 2nd-order DSUSY transformation.

The two-supermode multiplexing devices were then fabricated on a SOI wafer using complementary metal-oxide-semiconductor (CMOS)-compatible processes[11,49]. The devices were patterned by electron beam lithography (EBL) and etched via inductively coupled plasma (ICP) dry etching. A 1-μm-thick silica layer was deposited on top of the devices for cladding by plasma enhanced chemical vapor deposition (PECVD). Detailed fabrication procedures are provided in Methods and Supplementary Note 4. Figure 2d presents the optical micrograph of the fabricated $SM_1$ and $SM_2$ mode (de)multiplexer (left) and the zoomed-in scanning electron microscope (SEM) images of the areas of the waveguide array highlighted in the red and orange dashed boxes and represented by $H_{\text{DSUSY}}$ and $H_{\text{SM}}$, respectively (right). A tunable continuous-wave (CW) laser and a photodetector (PD) were employed to measure the transmission spectra of the fabricated devices. The experimental setup and the measurement methods are detailed in Methods and Supplementary Note 4. The measured transmission spectra of the $SM_1$ and $SM_2$ mode (de)multiplexer are shown in

Fig. 2e. In a wavelength range of 1500-1600 nm, the ILs are lower than 3.8 and 2.9 dB and the CT values are below -19.6 and -18.8 dB for the $SM_1$ and $SM_2$ channels, respectively (ILs of < 2.16 dB and CT of < -27 dB for both channels at 1550 nm). To validate the universality of the proposed method, we also fabricated the $SM_3$ and $SM_4$ mode (de)multiplexer and measured the transmission spectra of both channels (Fig. 2f). Over a 100-nm bandwidth spanning from 1500 to 1600 nm, the (de)multiplexer exhibits ILs less than 3.23 and 3.14 dB and CT smaller than -15.8 and -17.1 dB for the $SM_3$ and $SM_4$ channels, respectively (ILs of < 1.12 dB and CT of < -18 dB for both channels at 1550 nm). The discrepancies between theory and experiment could be attributed to scattering losses caused by waveguide sidewall roughness and imperfect couplings between waveguides due to partial filling of the gaps with a cladding material[50-52]. The device performance can be further improved by optimizing the fabrication process. Nonetheless, the experimental results provide convincing evidence that high-performance supermode multiplexing can be achieved by engineering an equidistant $n_{eff}$ distribution of supermodes and harnessing 2nd-order DSUSY transformations for accurate supermode evolution.

**Scalability of supermode multiplexing based on SUSY transformation**

To prove the scalability of our approach, we construct a four-channel supermode multiplexing system by simply cascading the previously introduced two-supermode (de)multiplexers. The 3D schematic of the device is shown in Fig. 3a, where the orange and purple dashed boxes highlight the multiplexing regions for the $SM_1$ and $SM_2$ modes and the $SM_3$ and $SM_4$ modes, respectively. The demultiplexing regions on the right are mirror-symmetric to their multiplexing counterparts. An enlarged view of the area encircled by the orange dashed box is provided in the right panel. It consists of four distinct areas (labeled as Regions I, II, III, and II', respectively), which are arranged in sequence to form the four-supermode multiplexer. Region I contains an adiabatic structure evolving from the DSUSY array ( $H_{\mathrm{DSUSY}}$ ) to the supermode array ( $H_{\mathrm{SM}}$ ), enabling the excitation of $SM_1$ and $SM_2$ supermodes through the isolated waveguide inputs. Then, it gradually transforms to the other SUSY structure described by $H'_{\mathrm{DSUSY}}$ in Region II, confining the $SM_1$ and $SM_2$ supermodes within the two inner waveguides of the array. Region III is composed of a pair of adiabatic couplers which launch the $TM_0$ modes into two isolated waveguides to excite the $SM_3$ and $SM_4$ supermodes (design details are given in Supplementary Note 5). Finally, Region II' possesses a mirrored structure with respect to Region II, allowing the uploaded supermodes to be transformed to those supported by the evanescently coupled waveguides ($H_{SM}$). Most importantly, since the $SM_1$ and $SM_2$ supermodes are well confined to the two inner waveguides of the array in Region III, injecting the $SM_3$ and $SM_4$ supermodes through the two outer waveguides does not affect the propagation of the already excited $SM_1$ and $SM_2$ supermodes, enabling high-fidelity generation and low-CT transmission of four orthogonal supermodes in the multiplexing region encircled by the purple dashed box (more details can be found in Supplementary Note 6).

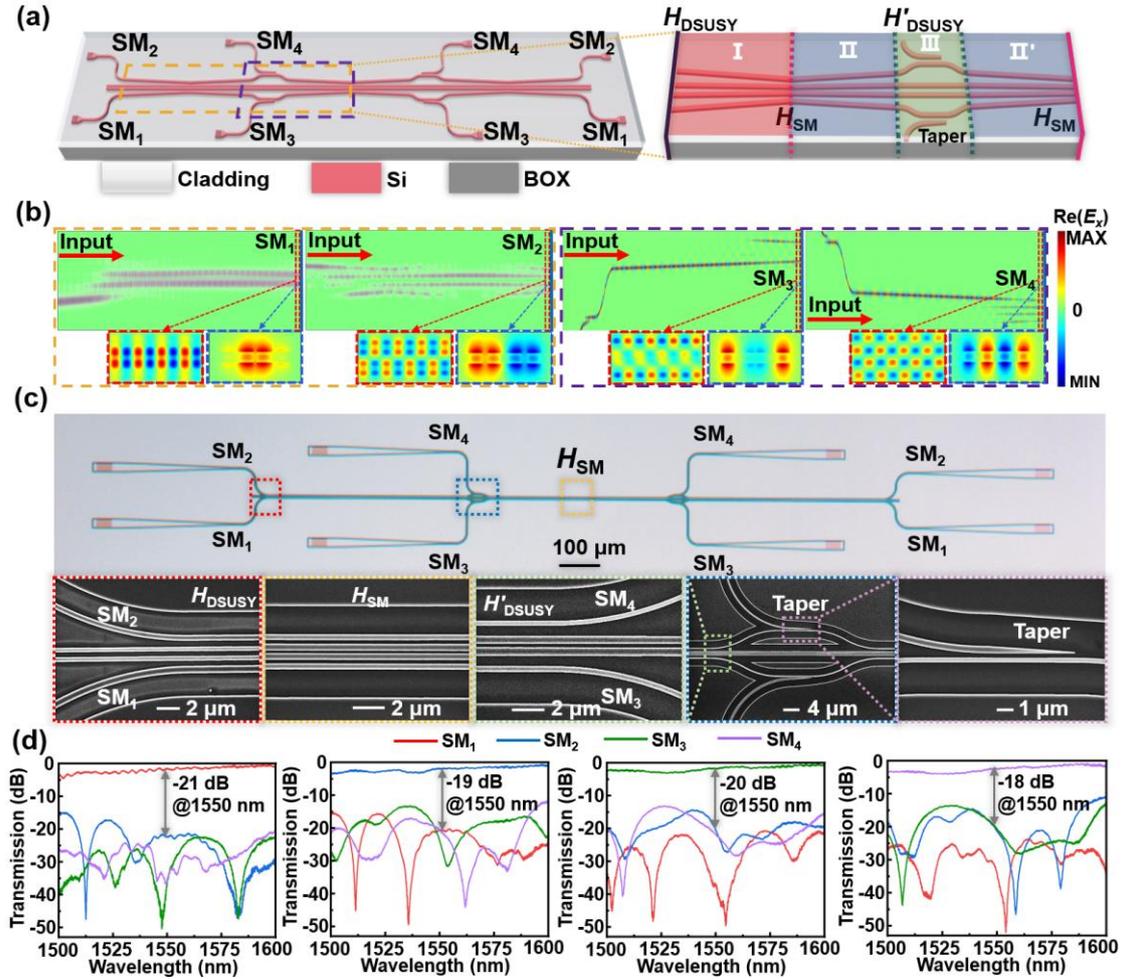

**Fig. 3| Scalability of supermode multiplexing enabled by 2nd-order DSUSY transformations.
a**, Schematic of the four-supermode multiplexing device (left) with an enlarged view of the $SM_1$ and $SM_2$ mode multiplexing region encircled by the orange dashed box (right). The $SM_3$ and $SM_4$ mode multiplexing region is highlighted by the purple dashed box. The demultiplexing structure is mirror-symmetric to the multiplexing one which contains four regions labeled I, II, III and II'. **b**, Simulated propagation profiles ( $\mathrm{Re}(E_x)$ ) for the $SM_1$-$SM_4$ supermodes when the light is launched from the corresponding input ports. The orange and purple dashed boxes indicate those areas encircled in **a**. The bottom insets show the top views (in the red dashed boxes) and cross-sectional views (in the blue dashed boxes) of the supermode profiles at the output end of the multiplexer. **c**, Optical micrograph of the fabricated device (top) and magnified SEM images of the DSUSY arrays, the supermode array and the adiabatic couplers (bottom). The correspondence between the areas shown in the figures are indicated by the colors of the dashed boxes surrounding them. **d,** Measured transmission spectra of the $SM_1$-$SM_4$ channels of the four-supermode (de)multiplexer.

To verify the feasibility of the proposed design, we performed 3D FDTD simulations to investigate the excitation and propagation of these supermodes in the device. Figure 3b shows the simulated propagation profiles ($\mathrm{Re}(E_x)$) for the $SM_1$-$SM_4$ supermodes in the $SM_1$ and $SM_2$ mode multiplexing region (encircled by the orange dashed boxes) and the $SM_3$ and $SM_4$ mode multiplexing region (encircled by the purple dashed boxes). The bottom insets present the top views (in the red dashed boxes) and cross-sectional views (in the blue dashed boxes) of the supermode profiles after passing the multiplexing regions. As one can see, all four supermodes can be selectively excited with high purity by launching the $TM_0$ modes into the corresponding input ports labeled $SM_1$-$SM_4$. The simulated transmission spectra of all channels were also extracted to evaluate the performance of the four-supermode (de)multiplexer (see Supplementary Note 7 for details). Across a wide wavelength range from 1500 to 1600 nm, the ILs are below 0.11, 0.19, 0.11 and 0.02 dB while the CT values are lower than -16.0, -16.3, -19.0 and -18.9 dB for the $SM_1$-$SM_4$ channels, respectively (ILs of < 0.08 dB and CT of < -22 dB for all channels at 1550 nm). These numerical results unequivocally prove that the DSUSY transformation enables scalable supermode multiplexing without compromising the performance of orthogonal channels such as bandwidths, ILs and CT. Moreover, the fabrication tolerance analysis reveals that the proposed device is highly resilient to fabrication errors, maintaining low ILs of < 2.1 dB for all channels at 1550 nm even with substantial deviations in waveguide width (-125 to 200 nm) or gap distance (-150 to 200 nm). Such a level of robustness is highly desirable for large-scale photonic integration (see Supplementary Note 8).

The four-channel supermode-division multiplexing device was then fabricated using the same process as described in the previous section. Figure 3c displays the optical micrograph of the fabricated device (top) and the zoomed-in SEM images of the DSUSY arrays, the supermode array and the adiabatic couplers (bottom). The coupled-waveguide structures corresponding to $H_{\text{DSUSY}}$, $H'_{\text{DSUSY}}$ and $H_{\text{SM}}$ are highlighted by the red, blue and orange dashed boxes respectively in the optical micrograph and shown in greater detail in the SEM images. The measured transmission spectra of all channels are depicted in Fig. 3d, where the (de)multiplexer shows ILs below 4.54, 3.55, 3.10 and 4.06 dB and CT less than -15, -12, -13 and -11 dB over a 100-nm bandwidth of 1500-1600 nm for the $SM_1$-$SM_4$ channels, respectively (ILs of < 2.48 dB and CT of < -18 dB for all channels at 1550 nm). These experimental results clearly demonstrate the scalability of the proposed DSUSY transformation method for broadband, low-loss and low-CT supermode multiplexing. Combined with supermode engineering featuring an equidistant $n_{\text{eff}}$ distribution, the system can be extended to more supermode channels as well as hybrid multiplexing schemes incorporating, for example, the supermode and polarization dimensions.

**High-speed data transmission experiment**

To demonstrate the practical applicability of the proposed device, we carried out a high-speed data transmission experiment based on the four-supermode multiplexing system by sequentially sending a high-baud-rate signal to each channel. The experimental setup and transceiver digital signal processing (DSP) algorithms are illustrated in Fig. 4a, b. A Nyquist-shaped 64-GBaud 16-quadrature amplitude modulation (16-QAM) signal was transmitted through each channel, corresponding to a per-channel data rate of 256 Gbit/s and a total capacity of 1.024 Tbit/s. At the receiver, a real-valued multiple-input multiple-output feedforward equalizer (MIMO-FFE) and maximum-likelihood sequence detection (MLSD) were employed to mitigate transmission impairments[53,54]. More detailed information about the experimental setup and the DSP implementation are provided in Methods and Supplementary Note 9. The measured optical spectra of the transmitted signals at different stages are presented in Fig. 4c. Figure 4d shows the BER performance of the four supermode channels, all below the KP4-FEC threshold of $2\times10^{-4}$. It implies that error-free transmission can be achieved by adopting standard FEC coding. The recovered 16-QAM constellations for all channels are plotted in Fig. 4e, showing well-resolved symbol clusters and therefore confirming high-fidelity transmission of the supermode-multiplexed signals. Taken together, these results prove the viability of harnessing supermodes as a new degree of freedom for high-capacity on-chip optical communications.

## Discussion

To conclude, we have proposed and experimentally demonstrated a systematic methodology that unlocks the supermode dimension as a new degree of freedom for light manipulation. The advance rests on two key innovations that together make supermode photonics practically viable.

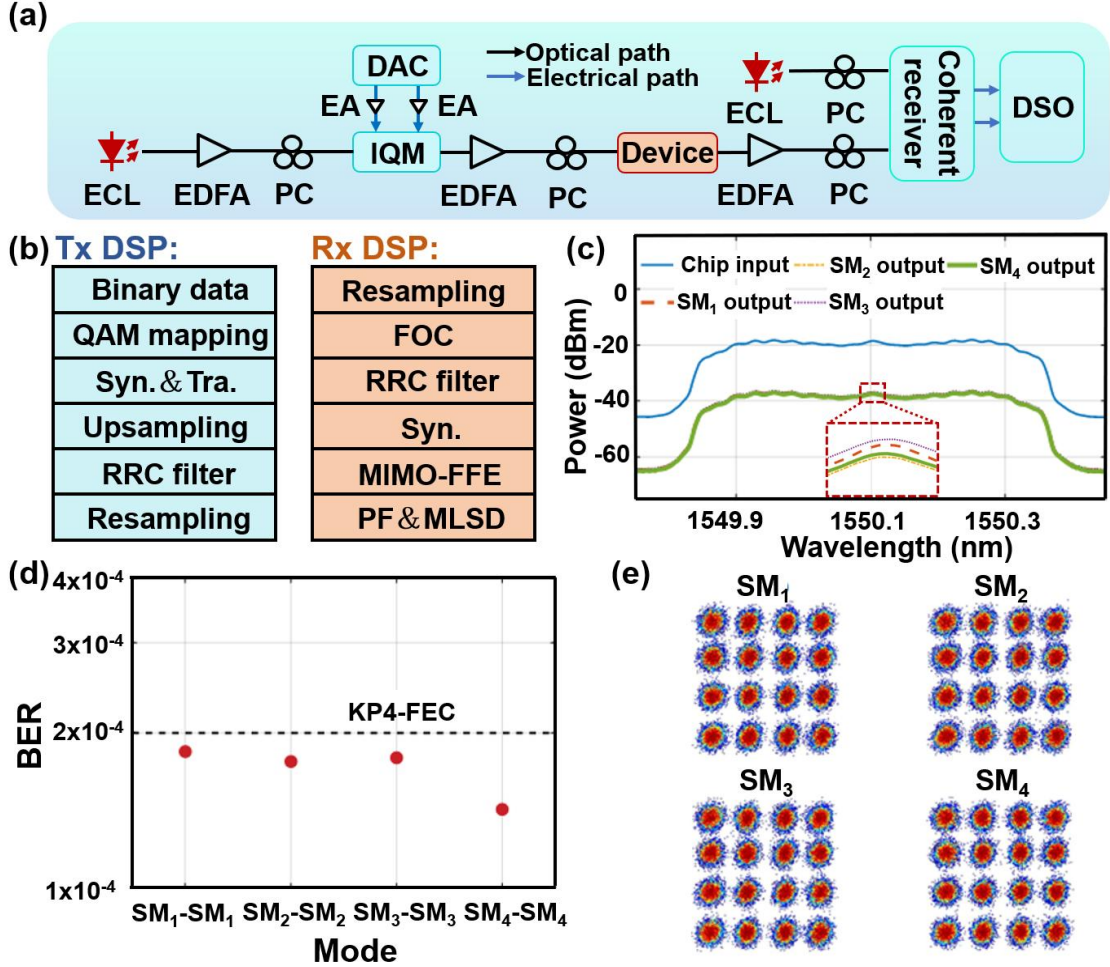

**Fig. 4 | High-speed data transmission experiment based on the four-supermode multiplexing system. a,** Experimental setup for the 256-Gbit/s data transmission per channel. The black and blue lines represent optical and electrical links, respectively. ECL: external cavity laser, EDFA: erbium-doped fiber amplifier, PC: polarization controller, DAC: digital-to-analog converter, EA: electrical amplifier, IQM: in-phase quadrature modulator, DSO: digital storage oscilloscope. **b,** DSP flowcharts for the transmitter (Tx) and receiver (Rx). **c,** Measured optical spectra of the 16-QAM signals before and after passing the chip. The inset provides a zoomed-in view of the spectra around 1550 nm, showing an uniform power distribution across the $SM_1$-$SM_4$ outputs. **d,** BERs for the four supermode channels. The horizontal dashed line denotes the KP4-FEC threshold of $2 \times 10^{-4}$. **e,** Recovered 16-QAM constellation diagrams for the $SM_1$-$SM_4$ channels.

First, by engineering the multi-well optical potential of a waveguide array, we obtained supermodes with an equidistant $n_{eff}$ distribution, which mitigates the degradation of mode orthogonality and the resulting increase in intermodal coupling under perturbations. Second, by leveraging the 2nd-order DSUSY transformation method[9], we developed an universal paradigm for the accurate excitation and extraction of arbitrary supermodes. To the best of our knowledge, this is the first work that combines these two technologies to achieve flexible and simultaneous control over multiple supermodes, paving the way towards integrated supermode photonics.

As proof-of-concept demonstrations, we experimentally implemented two- and four-supermode multiplexing systems to validate the feasibility and scalability of the proposed scheme. The fabricated devices show low ILs of < 4.54 dB and intermodal CT of < -11 dB for all channels in a wavelength band of 1500-1600 nm (ILs of < 2.48 dB and CT of < -18 dB at 1550 nm). Furthermore, a high-speed data transmission experiment was also conducted on the four-supermode (de)multiplexer with a 256-Gb/s 16-QAM signal per channel, achieving a total capacity of 1.024 Tb/s. The BERs fall below the KP4-FEC threshold for all channels, highlighting the potential of this approach for next-generation high-capacity on-chip optical interconnects.

Looking ahead, the findings in this work could inspire a wide range of future studies. For example, the design method can be applied to various integrated photonic platforms such as AlGaAs[55,56], thin-film lithium niobate[57,58] and silicon nitride[13,57] (see Supplementary Note 10). It can also be extended to the transverse electric (TE) polarization and enable the more advanced supermode- and polarization-division hybrid multiplexing devices (see Supplementary Note 11). High-efficiency conversion between conventional waveguide modes and supermodes will further facilitate the development of integrated supermode photonics and on-chip light manipulation (see Supplementary Note 12). At the system level, the total transmission capacity can be scaled as the product of the numbers of available supermodes, polarizations, wavelengths and guided modes per waveguide. Overall, our work marks a substantial advance in integrated supermode photonics, establishing a new paradigm in which supermodes serve as an additional degree of freedom of light for diverse signal processing applications, including on-chip optical communications[37,38], intelligent optical computing[2,6] and quantum information processing[39,40].

# Methods

**Sample fabrication**

The two- and four-supermode multiplexing devices were fabricated on a SOI wafer with a 220-nm-thick top silicon layer and a 3-μm-thick buried oxide layer. The wafer was sequentially cleaned in ultrasonic baths of acetone and isopropyl alcohol (IPA) for 5 min each, followed by $O_2$ plasma treatment. The device patterns were defined by EBL (Vistec EBPG 5200⁺) and transferred onto the top silicon layer via ICP dry etching (SPTS DRIE-I) with an etching depth of 220 nm. The above steps were repeated to fabricate the grating couplers (GCs), but this time with a shallower etching depth of 70 nm. Finally, a 1-μm-thick silica upper cladding layer was deposited on top of the devices by PECVD (Oxford Plasmalab System 100). The fabricated samples were inspected using an optical microscope and a SEM (Zeiss Ultra Plus). More details can be found in Supplementary Note 4.

**Optical Characterization**

In the experiments, a tunable CW laser (Santec TSL-770) and a PD (Santec MPM-210) were employed to characterize the fabricated devices. The polarization of light from the laser was first adjusted by a fiber PC (Thorlabs FPC032). Then, the TM-polarized light was coupled into and out of the chip through GCs. An optical power meter was used for calibration while the laser wavelength was swept from 1500 to 1600 nm to record the transmitted power at every wavelength using a PD positioned at the device output. More details can be found in Supplementary Note 4.

**High-speed transmission**

The experimental setup and the DSP flow for the 64-GBaud 16-QAM signal transmission are depicted in Fig. 4a, b. At the transmitter, binary data were mapped onto 16-QAM symbols, upsampled, and pulse-shaped using a root-raised-cosine (RRC) filter. After resampling, the waveforms were generated by a 100-GSa/s DAC (Micram DAC4) and subsequently amplified by an EA to drive a 35-GHz I/Q modulator. The modulator was biased at the null point to enable carrier-suppressed transmission. An ECL with a linewidth of 15 kHz and an outport power of 10 dBm was employed as the light source whose signals were amplified by an EDFA before modulation. The modulated signals were boosted by a second EDFA to compensate for the ILs before being launched into the supermode-division multiplexing device. At the receiver, the transmitted signals were optically pre-amplified, detected by a coherent receiver, and digitized by a 160-GSa/s oscilloscope (LeCroy 36Zi-A). The offline DSP included resampling, frequency offset compensation, matched filtering, synchronization and MIMO-FFE for linear distortion compensation. Finally, a post filter and MLSD were used to mitigate the MIMO-induced noise enhancement before BER evaluation. Please refer to Supplementary Note 9 for more details.

**Author contributions**

K. C. and L. S. conceived the idea. K. C., L. S., and Q. L. performed the theoretical derivation and numerical simulation. K. C. designed and fabricated the devices. K. C. and J. L. carried out the experimental measurements. Y. Z. and Y. L. assisted the fabrication and characterization. Z. L. and C. L. were involved in the discussions about theory and data analysis. K. C. and L. S. wrote the manuscript with input from all authors. L. S. and Y. S. supervised the project.

**Acknowledgements**

The work was supported in part by the National Key Research and Development Program of China (2023YFB2905503(C. L.)) and the National Natural Science Foundation of China (62475146(L. S.) and 62341508(Y. S.)). The authors thank the Center for Advanced Electronic Materials and Devices

(AEMD) of Shanghai Jiao Tong University (SJTU) for the support in device fabrication.

**Competing interests**

The authors declare no competing interests.

**Data availability**

The data that support the findings of this study are provided in the Supplementary Information/Source Data file. Source data are provided with this paper.